\documentclass[aps,twocolumn,prl,tightenlines,floatfix,superscriptaddress,longbibliography]{revtex4-1}

\usepackage[breaklinks=true]{hyperref}
\usepackage{graphicx}
\usepackage[english]{babel}
\usepackage{amsmath}
\usepackage{amsfonts}
\usepackage{amssymb}
\usepackage{times}

\renewcommand{\d}{\textrm{d}}
\newcommand{\be}{\begin{equation}}
\newcommand{\ee}{\end{equation}}
\newcommand{\e}{\mathrm{e}}

\begin{document}

\title{Exotic superfluidity and pairing phenomena in atomic Fermi
  gases in mixed dimensions}

\author{ Leifeng Zhang }
\author{Yanming Che }
\affiliation{Department of Physics and Zhejiang Institute of Modern
  Physics, Zhejiang University, Hangzhou, Zhejiang 310027, China}
\affiliation{Synergetic Innovation Center of Quantum Information and
  Quantum Physics, Hefei, Anhui 230026, China} 

\author{Jibiao Wang }
\affiliation{Department of Physics and Zhejiang Institute of Modern
  Physics, Zhejiang University, Hangzhou, Zhejiang 310027, China}
\affiliation{Synergetic Innovation Center of Quantum Information and
  Quantum Physics, Hefei, Anhui 230026, China} 
\affiliation{TianQin Research Center \& School of Physics and Astronomy, Sun Yat-Sen University (Zhuhai Campus), Zhuhai, Guangdong 519082, China} 

\author{Qijin Chen}
\email[Corresponding author: ]{qchen@uchicago.edu}
\affiliation{Department of Physics and Zhejiang Institute of Modern
  Physics, Zhejiang University, Hangzhou, Zhejiang 310027, China}
\affiliation{Synergetic Innovation Center of Quantum Information and
  Quantum Physics, Hefei, Anhui 230026, China} 
\affiliation{James Franck Institute, University of Chicago, Chicago, Illinois
  60637, USA}

\date{\today}

\begin{abstract}
  Atomic Fermi gases have been an ideal platform for simulating
  conventional and engineering exotic physical systems owing to their
  multiple tunable control parameters.  Here we investigate the
  effects of mixed dimensionality on the superfluid and pairing
  phenomena of a two-component ultracold atomic Fermi gas with a
  short-range pairing interaction, while one component is confined on
  a one-dimensional (1D) optical lattice whereas the other is in a
  homogeneous 3D continuum.  We study the phase diagram and the
  pseudogap phenomena throughout the entire BCS-BEC crossover, using a
  pairing fluctuation theory. We find that the effective
  dimensionality of the non-interacting lattice component can evolve
  from quasi-3D to quasi-1D, leading to strong Fermi surface
  mismatch. Upon pairing, the system becomes effectively quasi-two
  dimensional in the BEC regime.  The behavior of $T_c$ bears
  similarity to that of a regular 3D population imbalanced Fermi gas,
  but with a more drastic departure from the regular 3D balanced case,
  featuring both intermediate temperature superfluidity and possible
  pair density wave ground state. Unlike a simple 1D optical lattice
  case, $T_c$ in the mixed dimensions has a constant BEC asymptote.
\end{abstract}


\maketitle

Ultracold atomic gases have been under active investigation in the
past decades with their remarkable tunability in terms of interaction,
population and mass imbalance \cite{Review,Bloch_RMP}, and so on. They
have provided an ideal platform for simulating existing and
engineering exotic physical systems. Therefore, besides the atomic and
molecular physics community, they have attracted a lot of attentions
from other fields of physics, including condensed matter, nuclear
matter, color superconductivity, etc.  In particular, they can be put
in an optical lattice, \cite{Bloch05NP} with variable lattice depth
and spacing, which controls the hopping integral between neighboring
lattice sites. This provides an exciting opportunity for studying
exotic many-body phenomena caused by tuning the dimensionality
\cite{CWu15PRL,CWu16}. Among others, of great interest are fermion
pairing and related superfluid phenomena in mixed dimensions
\cite{Nishida08PRL,Nishida10PRA}. 

Recently, Lamporesi \textit{et al.}  \cite{Lamporesi10PRL} has
successfully obtained a mixed-dimensional system with a Bose-Bose
mixture of $^{41}$K--$^{87}$Rb using a species-selective
one-dimensional (1D) optical lattice technique; only $^{41}$K atoms
feel the lattice potential, leaving $^{87}$Rb atoms moving freely in
the 3D continuum. They observed a series of resonances in the mixed
dimensions. Motivated by this experiment, there have been theoretical
investigations of the BCS--Bose-Einstein condensation (BEC) crossover
in Fermi gases in mixed dimensions. Iskin and coworkers
\cite{Iskin10PRA} investigated the phase diagrams of equal population
fermion mixtures at zero temperature $T$ using a strict mean-field
approach and found the phase diagram in some ways similar to the Sarma
state in a usual 3D Fermi gas with a population imbalance. In order to
address real experiments, studies of phase diagrams at finite
temperatures are necessary. However, so far only preliminary study of
very limited cases at the finite temperature has be reported in the
literature \cite{XYang11EPJB}.

In this paper, we explore systematically the effects of mixed
dimensionality on the pairing and superfluidity at finite temperatures
in two-component ultracold atomic Fermi gases. Due to the high
complexity caused by multiple tunable parameters, here we restrict
ourselves to the population balanced case with equal masses, in order
to single out the effects of the dimensionality mismatch. For the same
reason, here we will not consider possible
Fulde-Ferrell-Larkin-Ovchinnikov (FFLO) states \cite{FF,LO} and phase
separation, leaving them to future studies.

We shall consider the same dimensionality setting as in the experiment
of Ref.~\cite{Lamporesi10PRL}, and refer to the lattice and 3D
continuum components as spin up and spin down, respectively.

To address the finite temperature effects, we use an existing pairing
fluctuation theory, which includes self-consistently the contributions
of finite momentum pairs \cite{Chen2,Review}, and has been applied to
address multiple experiments \cite{FrontPhys}.  We study the behavior
of the superfluid transition temperatures $T_c$ as a function of
interaction strength throughout the entire BCS-BEC crossover with a
varying optical lattice spacing $d$ and tunneling matrix element
$t$. We find that this non-polarized mixed-dimensional finite $T$
result share features in common with a polarized Fermi gas in a simple
3D continuum \cite{Chien06,Guo2009PRA}.  Our results show that the
closest match between the Fermi surfaces of the two pairing components
occurs near $t/E_F=1$ and $k_Fd=1$. (Here the Fermi momentum $k_F$ and
energy $E_F\equiv \hbar^2k_F^2/(2m)$ are defined via the 3D
component). Deviation from these parameters lead to drastic Fermi
surface mismatch, and the resulting phase diagrams can become quite
different from their counterpart of the polarized Fermi gases in
regular 3D continuum.  For a large range of parameters of $d$ and $t$,
the superfluid phase in the unitary regime may extend all the way down
to $T=0$, allowing a zero $T$ superfluid ground state. This is
distinct from the population imbalanced Fermi gas case in regular 3D
continuum, where an arbitrarily small but finite population imbalance
is sufficient to destroy superfluidity at zero $T$ at unitarity.

We use the same formalism as given in Ref.~\cite{Chien06}, which is
now adapted for the mixed dimensions.  For the lattice dimension, we
use a one-band nearest-neighbor tight-bind lattice model, and thus the
dispersions for the two components are
$\xi_{\mathbf{k}\uparrow}=\mathbf{k}_\parallel^2/2m+2t[1-\cos(k_zd)]-\mu_\uparrow$
and $\xi_{\mathbf{k}\downarrow}=\mathbf{k}^2/2m-\mu_\downarrow$. Here
$\mathbf{k}_\parallel\equiv(k_x,k_y)$, where $\mu_\sigma$ (with
$\sigma=\uparrow,\downarrow$) are the fermionic chemical potentials.
The one-band assumption is appropriate when the lattice band gap is
large compared to Fermi energy $E_F$, which may be realized
experimentally via a large confining trapping frequency
$\omega_z \gg E_F$. Indeed, similar one-band models have been used
throughout condensed matter theory studies, e.g., in various high
$T_c$ superconductivity theories and negative-$U$ Hubbard models
\cite{Micnas}. (In particular, the typical condensation energy per
particle is much less than $E_F$, even at unitarity, and thus not
enough to compensate the energy cost for a lattice fermion to
occupy the excited bands.) As usual, we consider an $s$-wave short
range pairing interaction.  The bare fermion Green's functions are
given by $G^{-1}_{0\sigma}(K)=i\omega_n-\xi_{\mathbf{k}\sigma}$. We
refer the readers to Ref.~\cite{Chien06} for convention on the
four-vector notations and Matsubara frequencies.

Following Ref.~\cite{Chien06}, the noncondensed pair contributions to
the self-energy in the superfluid phase can be well approximated as
$\Sigma_{pg,\sigma}(K)\approx [\sum_Q
t_{pg}(Q)]G_{0\bar{\sigma}}(-K)$,
in the same form as the superconducting self energy $\Delta_{sc}(K)$,
after defining a pseudogap parameter $\Delta_{pg}$ via
$\Delta_{pg}^2\equiv {-\sum_Q t_{pg}(Q)}$, where $t(Q)$ is the pairing
$T$ matrix, and $\bar{\sigma}=-\sigma$. Then we obtain a total
self-energy in the BCS form
$\Sigma_\sigma(K)=-\Delta^2G_{0\bar{\sigma}}(-K)$, where
$\Delta^2=\Delta_{sc}^2+\Delta_{pg}^2$. This immediately leads to the
full Green's functions
\begin{eqnarray}
  G_\sigma(K)&=&\frac{u_\mathbf{k}^2}{i\omega_n-E_{\mathbf{k}\sigma}}
             +\frac{v_\mathbf{k}^2}{i\omega_n+E_{\mathbf{k}\bar{\sigma}}},\quad
             |k_z|<\frac{\pi}{d}\nonumber\\
G_\downarrow(K)&=&\frac{1}{i\omega_n-\xi_{\mathbf{k}\downarrow}}, \quad
|k_z|>\frac{\pi}{d}\,
  \label{eq:GF}
\end{eqnarray}
where $u_\mathbf{k}^2=(1+\xi_\mathbf{k}/E_\mathbf{k})/2,
v_\mathbf{k}^2=(1-\xi_\mathbf{k}/E_\mathbf{k})/2$,
$E_\mathbf{k}=\sqrt{\xi_\mathbf{k}^2+\Delta^2}$, and
$E_{\mathbf{k}\sigma}=E_\mathbf{k}+\zeta_{\mathbf{k}\sigma}$,
$\xi_\mathbf{k}=(\xi_{\mathbf{k}\uparrow}+\xi_{\mathbf{k}\downarrow})/2$,
$\zeta_{\mathbf{k}\sigma}=(\xi_{\mathbf{k}\sigma}-\xi_{\mathbf{k}\bar{\sigma}})/2$.
Note that $k_{z\uparrow}$ has been restricted to within the first
Brillouin zone, $[-\pi/d,\pi/d]$, due to the lattice periodicity.

With $n_\sigma=\Sigma_KG_\sigma(K)$, we obtain the total atomic number density
$n=n_\uparrow+n_\downarrow$ and the difference $\delta
n=n_\uparrow-n_\downarrow = 0$ as
\begin{eqnarray}
  n&=&2\sum_\mathbf{k}\left[
      v_\mathbf{k}^2
      +\bar{f}(E_\mathbf{k})\frac{\xi_\mathbf{k}}{E_\mathbf{k}}
    \right]
    +\!\!\!\!\sum_{|k_z|>{\pi}/{d}}\!\!\!\! f(\xi_{\mathbf{k}\downarrow})\,,
  \label{eq:ntot}
\\
0&=&\sum_\mathbf{k}\left[
    f(E_{\mathbf{k}\uparrow})-f(E_{\mathbf{k}\downarrow})
  \right]
  -\!\!\!\!\sum_{|k_z|>{\pi}/{d}}\!\!\!\! f(\xi_{\mathbf{k}\downarrow})\,,
  \label{eq:dn}
\end{eqnarray}
where $f(x)$ is the Fermi distribution function, and the average
$\bar{f}(x)\equiv{\sum_\sigma f(x+\zeta_{\mathbf{k}\sigma})/2}$. In
contrast to the counterparts in the pure 3D continuum case, there is
an extra term of the 3D component in these equations, which has been
overlooked in Refs.~\cite{Iskin10PRA,XYang11EPJB}. When the Fermi
energy $E_F$ is lower than the lattice bandwidth $4t$, its
contribution is small. However, its contribution will become large
when $t$ is small, which is relevant to most 1D optical lattices in
experiment as of today.

After Nishida and Tan \cite{Nishida08PRL}, we use an effective
$s$-wave scattering length $a$ in the presence of the mixed
dimensionality to characterize the interaction strength between
fermions, via the Lippmann-Schwinger relation
$g^{-1}=m/4\pi a-\sum_\mathbf{k}1/2\epsilon_\mathbf{k}$. Here
$\epsilon_\mathbf{k}=(\epsilon_{\mathbf{k}\uparrow} +
\epsilon_{\mathbf{k}\downarrow})/2$,
with
$\epsilon_{\mathbf{k}\sigma}=\xi_{\mathbf{k}\sigma}+\mu_\sigma$. Note
that this scattering length in necessarily different from that defined
in ordinary 3D or 2D continuum, and is relevant to the actual
scattering length in the presence of the optical lattice, via, e.g.,
the binding energy $\epsilon_B = \hbar^2/2m_r a^2$ in the BEC
regime. In this way, the divergance of the scattering length $a$
corresponds to the threshold interaction strength $g_c$ for two
fermions to form a zero binding energy bound state in the mixed
dimensions, and where the actual $s$-wave scattering phase shift is
$\pi/2$, i.e., the unitary scattering.  In the superfluid state, the
Thouless criterion leads to the gap equation
\begin{equation}
  \frac{m}{4\pi a}=\sum_\mathbf{k}\left[
    \frac{1}{2\epsilon_\mathbf{k}}-\frac{1-2\bar{f}(E_\mathbf{k})}{2E_\mathbf{k}}
  \right].
\label{eq:gap}
\end{equation}
To better reflect the lattice contribution, we may deduce an effective
mass, $m_{eff}$, from the trace of the inverse mass tensor,
$\dfrac{1}{m_{eff}} = \dfrac{5}{6m}+\dfrac{1}{3}td^2 $, and then
define an effective scattering length $a_{eff}$ such that
$\dfrac{m}{4\pi a} = \dfrac{m_{eff}}{4\pi a_{eff}}$, or
$\dfrac{1}{k_Fa_{eff}} = \dfrac{1}{k_Fa}\dfrac{m}{m_{eff}}$. In
comparison with scattering length $a$, $a_{eff}$ reflects better the
actual scattering length that can be measured experimentally
\cite{Lamporesi10PRL}.

The inverse $T$ matrix can be expanded as
$t_{pg}^{-1}(Q)\approx{Z_1 (i\Omega_l)^2 +
  Z(i\Omega_l-{\Omega}_\mathbf{q})}$
in the superfluid phase \cite{Review}, where
${\Omega}_\mathbf{q}=q_\parallel^2/2M_\parallel^*+q_z^2/2M_z^*$, with
$M_\parallel^*$ and $M_z^*$ denoting the anisotropic effective pair
masses in the long wavelength limit.
Here we align the optical lattice in the $\hat{z}$ direction, so that
$\mathbf{q}_\parallel$ and $q_z$ are the in-plane and out-of-plane
pair momenta, respectively. The coefficients $Z$, $Z_1$,
$1/M_\parallel^*$ and $1/M_z^*$ can be computed from straightforward
Taylor expansion of the pair susceptibility at
$(\Omega,\mathbf{q}) = 0$.  It follows that the pseudogap contribution
\begin{equation}
  \Delta_{pg}^{2}=\sum_{\textbf{q}}\frac{b(\tilde{\Omega}_{\textbf{q}})}
  {Z\sqrt{1+4\dfrac{Z_{1}}{Z}  {\Omega}_\mathbf{q}}}\,,
  \label{eq:pg}
\end{equation}
where $b(x)$ is the Bose distribution function and
$\tilde{\Omega}_{\textbf{q}}=Z\{\sqrt{1+4Z_{1}
  {\Omega}_{\textbf{q}}/Z}-1\}/2Z_{1}$
is the pair dispersion. When $Z_1 \ll Z$, we have
$\tilde{\Omega}_{\textbf{q}} \approx {\Omega}_{\textbf{q}}$. The
integral over $q_z$ should be restricted to the first Brillouin Zone,
$|q_z|<\pi/d$, since in principle, $\Omega_\mathbf{q}$ will acquire
periodicity in $q_z$ as determined by the optical lattice. To a good
approximation, one may write
${\Omega}_\mathbf{q}=q_\parallel^2/2M_\parallel^*+2t_B[1-\cos(q_zd)]$,
with $t_B = 1/(2M^*_z d^2)$. We have checked numerically that using
this band dispersion would only cause slight quantitative difference
in $T_c$, as one can see from Supplementary Fig.~S2.

Equations~(\ref{eq:ntot})--(\ref{eq:pg}) form a closed set, which will
be used to solve for the superfluid transition temperature $T_c$ (and
the pseudogap $\Delta_{pg}$ and chemical potentials at $T_c$), by
setting the order parameter $\Delta_{sc}=0$. In the superfluid phase,
they can be used to solve for various gap parameters as well as
corresponding chemical potentials as a function of $T$.

The deep BEC regime can be worked out analytically, where everything
is small compared with $|\mu|$. Equation (\ref{eq:dn}) drops out, and
we obtain
\begin{subequations}
\begin{eqnarray}
\mu &=& - \left[\frac{1}{4m}\left(\frac{\pi}{d}\right)^2 + 2t\right] \e ^{\frac{d}{a}-2C}\,, \\
n &=& -\dfrac{m\Delta^2}{4\pi \mu d}, \qquad \Delta = \sqrt{\dfrac{-4\pi\mu d n}{m}}\,,\\
\dfrac{1}{2M^*_\parallel} &=& \dfrac{1}{2M^*_z} = \dfrac{1}{4m} 
\,,
\label{eq:Mpair}
\end{eqnarray}
\end{subequations}
where the constant
$\displaystyle C = \dfrac{d}{\pi}\int_0^{\pi/d} \dfrac{k_z^2+2m t d
  k_z \sin (k_zd)}{k_z^2+4m t [1-\cos (k_zd)]} \d k_z$
only depends on $t,d,m$, and takes the value between 0.7 and 1.0. It
is interesting to note that, from Eq.~(\ref{eq:Mpair}), the effective
pair mass $M^*$ approaches $2m$ in both in-plane and out-of-plane
directions. As a consequence, $T_c$ for all cases will approach
roughly the same BEC asymptote, which depends weakly on the lattice
constant $d$. This should be contrasted to the counterpart case in
which the $z$ direction is a lattice for both spins so that $1/2M^*_z$
(and hence $T_c$) shall decrease with increasing pairing strength in
the BEC limit.

Upon our solutions, we shall also enforce a positive definite
compressibility \cite{PWY05}, which has been shown to be equivalent to
the following condition \cite{Stability}:
\begin{equation}
  \left.\frac{\partial^2\Omega_S}{\partial \Delta^2}\right|_{\mu_\uparrow,\mu_\downarrow} \!\!\!\! =
  2\sum_\mathbf{k}\frac{\Delta^2}{E_\mathbf{k}^2}\left[
    \frac{1-2\bar{f}(E_\mathbf{k})}{2E_\mathbf{k}}+\bar{f}^\prime(E_\mathbf{k})
  \right]>0\,,
  \label{eq:stability}
\end{equation}
where $\bar{f}'(x) = \d \bar{f}(x)/\d x$, and $\Omega_S$ is the
thermodynamic potential, whose formal expression can be found in
Ref.~\cite{wang13pra}.  Phase separation may occur when this stability
condition is not satisfied.

\begin{figure}
  \centerline{\includegraphics[clip,width=3.in] {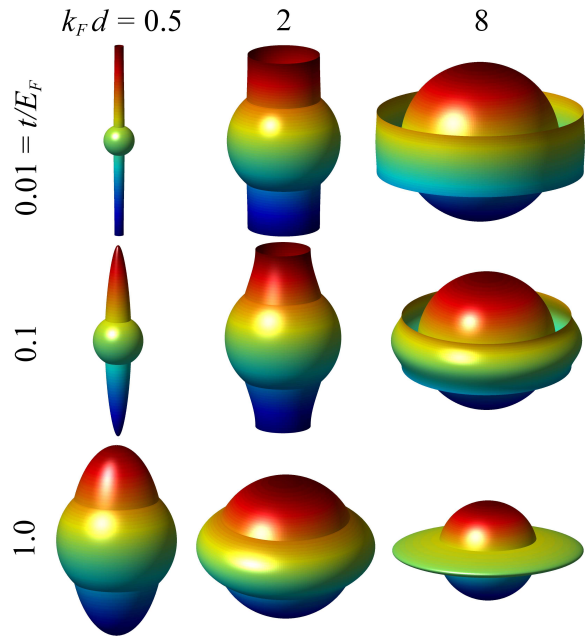}}
  \caption{Evolution of the Fermi surface of the
    lattice component as a function of $t$ and $d$, as compared with
    that of the 3D component (represented by the sphere). The Fermi
    surface is more like quasi-1D for small $d$ and quasi-2D for large
    $d$. }
\label{fig:FS}
\end{figure}

Before we present our solutions on the phase diagrams, let's first
study the Fermi surface mismatch in the noninteracting limit. In
Fig.~\ref{fig:FS}, we show how the Fermi surface of the lattice
component evolves as a function of $t$ and $d$, as compared with the
3D component, which is represented by the sphere. The closest match
occurs near $t/E_F = 1$ and $k_Fd=1$ (not shown). For fixed $t$, the
Fermi surface of the lattice component evolves from an elongated cigar
shape (quasi 1D) to a pan cake or disc (quasi 2D), as $d$
increases. On the other hand, for fixed $d$, the Fermi surface may
change from a pan cake (quasi 2D) to a cigar or a long cylinder (quasi
1D), as $t$ decreases. This can be readily understood. When $t$ is
small, it is more energetically favorable to populate on the $k_z$
quantum levels than the in-plane $k_\parallel$ levels. However, if $d$
is large, the first band $|k_z|<\pi/d$ becomes quickly filled so that
fermions have to accumulate in high $k_\parallel$ levels, leading to a
disc-like Fermi surface for small $t$ and large $d$. It is this case
which is mostly relevant to real experimental configurations, which
may be expected to satisfy $td^2 < 1/2m$.

Figures \ref{fig:FS} reveals that large Fermi surface mismatch may
occur for large and small $(d,t)$.  We shall now see this mismatch
effect at the mean-field level first.

\begin{figure}
  \centerline{\includegraphics[clip,width=3.2in] {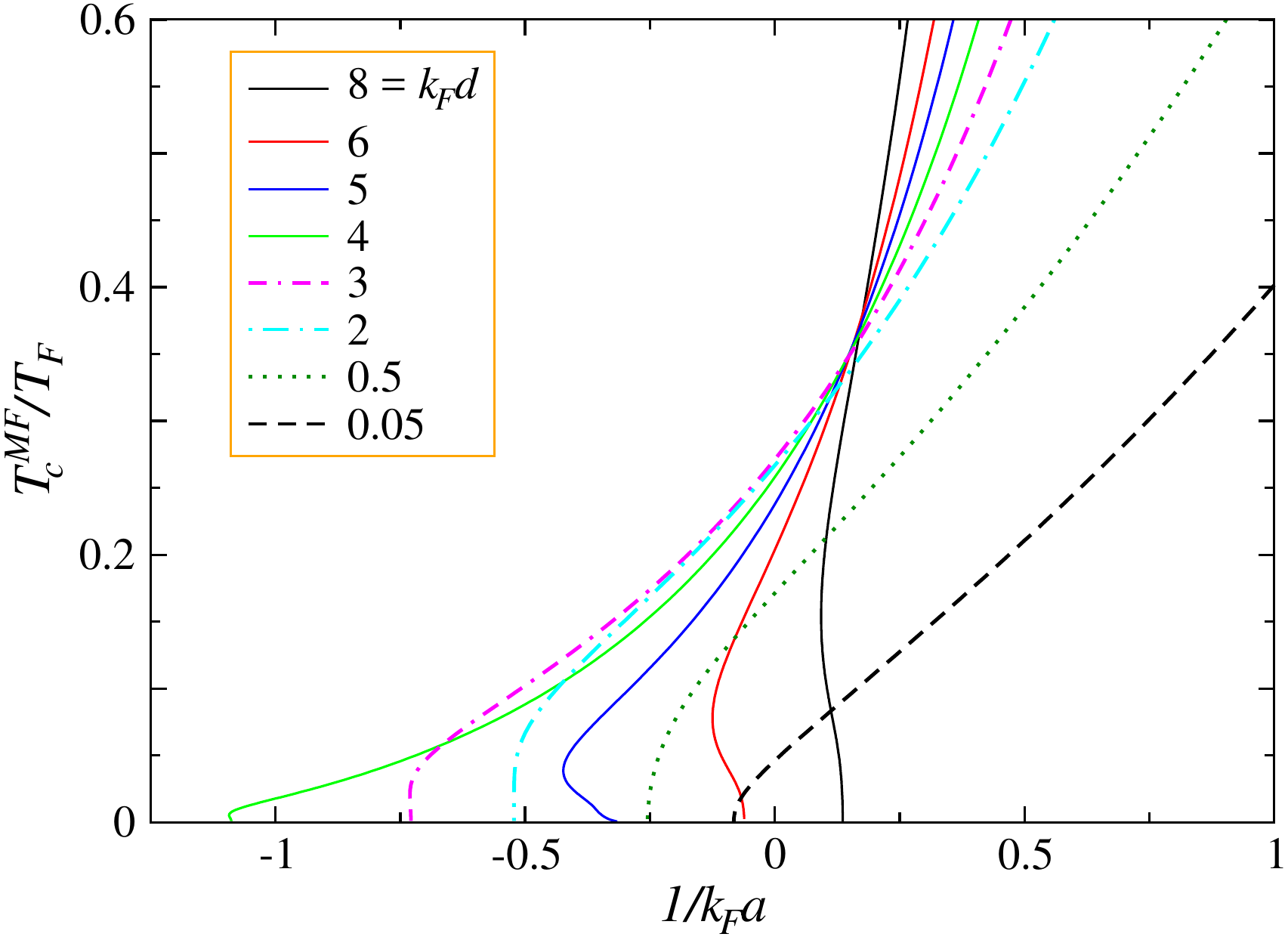}}
  \caption{Mean-field solution of $T_c^{MF}$ as a function of $1/k_Fa$
    for different $d$ with $t/E_F=0.05$. Intermediate temperature
    superfluidity occurs for $k_Fd = 5,6$ and 8.}
\label{fig:MF_Tc-d}
\end{figure}

Mean-field solutions can be obtained by solving
Eqs.~(\ref{eq:ntot})--(\ref{eq:gap}), assuming that the gap is the
order parameter. Shown in Fig.~\ref{fig:MF_Tc-d} are a series
mean-field $T_c$ curves as a function of $1/k_Fa$ with different $d$
and fixed $t/E_F =0.05$. For this small value of $t$, the best Fermi
surface match occurs near $k_Fd = 4$, in which case, the $T_c^{MF}$
curves to the left most into the BCS regime. As $d$ increases (solid
lines) or decreases (dashed lines), the curves, esp. their low $T$
thresholds, move towards stronger coupling. In other words, these
large or small $d$ values have stronger pair breaking effects at low
$T$ so that stronger pairing strength is needed to achieve
pairing. For $k_Fd>4$, there is clear evidence for intermediate
temperature superfluidity, as found in conventional population
imbalanced Fermi gases in a simple 3D continuum \cite{Chien06}.

We now proceed and present our main result with pairing fluctuation
effects included. While the $(t/E_F, k_Fd)=(1,1)$ possess the highest
Fermi surface match, such a large $t$ value is hard to realized
experimentally.  As a reference, we present this case in Supplementary
Fig.~S1.  Here we present in Fig.~\ref{fig:Tc-d} a more realistic case
of $t/E_F = 0.05$, and plot $T_c$ as a function of $1/k_Fa_{eff}$ for
a series of $d$ from large to small. For this case, the best Fermi
surface match occurs near $k_Fd = 4$ (See Fig.~\ref{fig:FS}), for
which the $T_c$ curve extends the deepest into the BCS regime, similar
to the mean-field case. As $d$ becomes smaller (dashed lines), the
threshold for the $T_c$ curve moves to the right, similar to the
mean-field result, and the $T_c$ values are suppressed at the same
time. For $k_Fd =0.25$ and smaller (0.05), $T_c$ is pinched and split
into two parts at intermediate coupling strength, in the regime around
$\mu = 0$, exhibiting a re-entrant superfluidity.

\begin{figure}
  \centerline{\includegraphics[clip,width=3.2in] {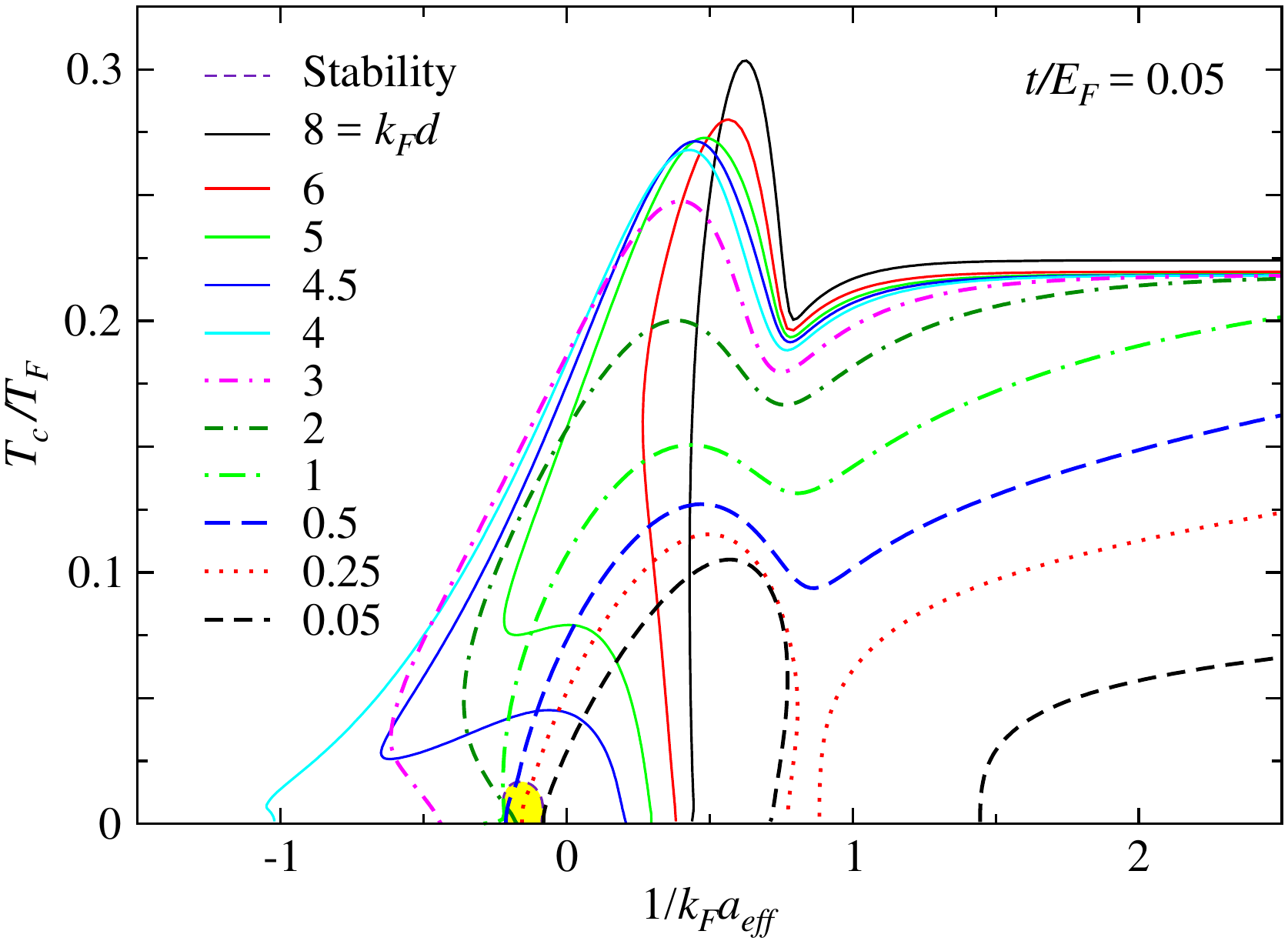}}
  \caption{Behavior of $T_c$ as functions of
    $1/k_Fa_{eff}$ at fixed $t/E_F = 0.05$, but for different value of
    $k_Fd$ from 8 to 0.05. The $T_c$ solution in shaded regions is
    unstable against phase separation.}
\label{fig:Tc-d}
\end{figure}

\begin{figure}
  \centerline{\includegraphics[clip,width=3.2in]{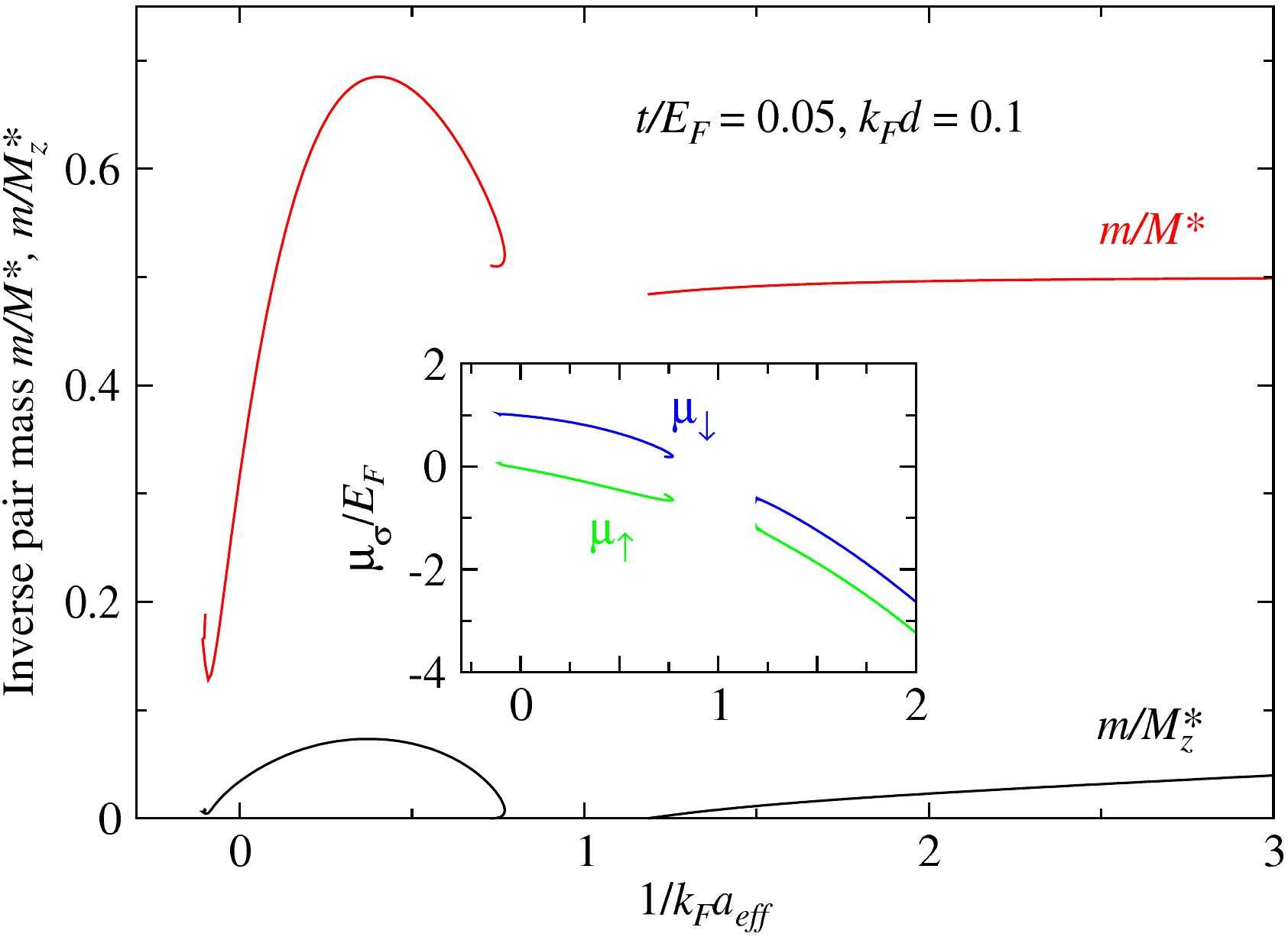}}
  \caption{Behavior of the in-plane (red) and out-of-plane (black)
    components of the inverse pair masses (main panel) and chemical
    potentials (inset, $\mu_\uparrow$ and $\mu_\downarrow$, as
    labeled) as a function of $1/k_Fa_{eff}$, for $t/E_F = 0.05$ and
    $k_Fd = 0.1$. The sign changes in $m/M_z^*$ lead to pair density
    wave ground state in between, exhibiting reentrant
    superfluidty. Here $M^*\equiv M^*_\parallel$.}
\label{fig:B}
\end{figure}

Such a re-entrant $T_c$ behavior was previously seen in the crossover
regime in a dipolar Fermi gas \cite{ChePolar}. This is a regime which
interpolates the BCS and the BEC regimes, where real space pairs start
to emerge as well defined composite particles while the inter-pair
repulsive interaction is very strong. For the present case, as can be
seen from Fig.~\ref{fig:FS}, the highly elongated quasi-1D Fermi
surface of the lattice component for small $d$ causes a large Fermi
surface mismatch. This mismatch then strongly suppresses the mobility
of the pairs in the $\hat{z}$ direction, leading to possible Wigner
crystallization of the pairs, and hence a pair density wave (PDW)
ground state without superfluidity. The Wigner crystallization is
signaled by a sign change of the effective pair mass at zero momentum,
as shown in Fig.~\ref{fig:B}. In the PDW state, the pair dispersion
would reach its minimum at a finite momentum. Such a potential energy
driven PDW state should not be confused with an FFLO states. The plot
of $\mu_\sigma$ in the inset of Fig.~\ref{fig:B} reveals that the
chemical potential for the lattice component is very small in size in
the BCS regime for this small value of $k_Fd$.
 
\begin{figure}[b]
  \centerline{\includegraphics[clip,width=3.2in] {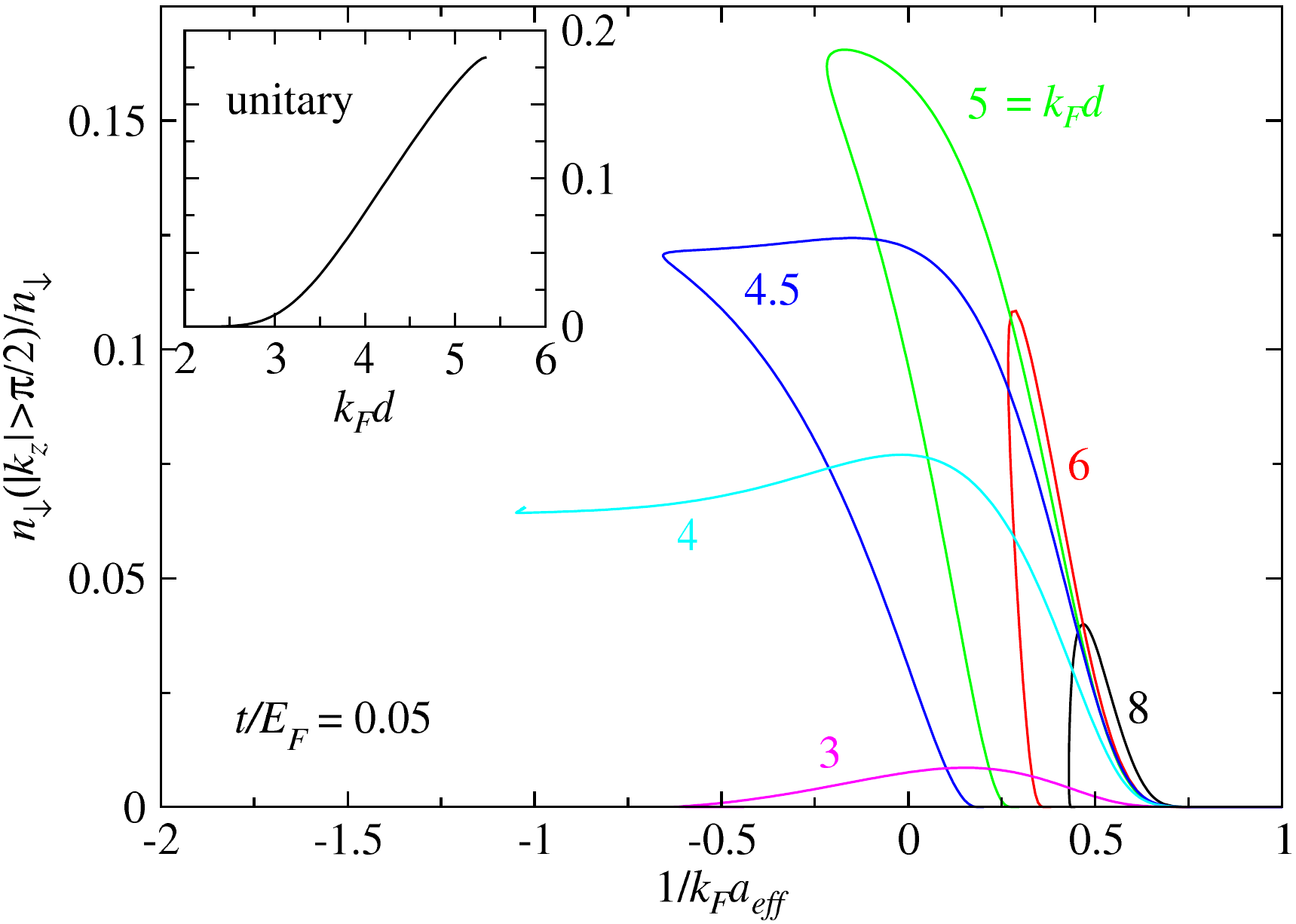}}
  \caption{Fraction of the 3D component outside the first Brillouin
    zone along the $T_c$ curves with $t/E_F = 0.1$ for different
    values of $k_Fd$. The inset plots the fraction at $T_c$ as a
    function of $k_Fd$ at unitarity. The fraction increases with
    $k_Fd$ but vanishes for all cases in the BEC regime.}
\label{fig:fraction}
\end{figure}

On the other hand, as $k_Fd$ increases from 4 (solid lines in
Fig.~\ref{fig:Tc-d}), the lattice Fermi surface becomes a disc, and
the lattice component becomes more 2D like. While this also leads to a
large Fermi surface mismatch, its damage can be substantially
alleviated when the pairing interaction becomes strong, since pairing
effectively prevents the 3D component from occupying large $|k_z|$
states, making it a quasi-2D system as well. To see this, we plot the
fraction of the 3D component with $|k_z|>\pi/d$. This effect is
manifested in Fig.~\ref{fig:fraction}, where we plot this fraction as
a function of $1/k_Fa_{eff}$ for different $d$'s, calculated along the
$T_c$ curves. It is obvious that the fraction increases with $d$ for
given $1/k_Fa_{eff}$. (Shown in the inset is a continuous curve as a
function of $d$ for the unitary case). A large fraction results from a
large Fermi surface mismatch. As $1/k_Fa_{eff}$ progresses into the BEC
regime, this fraction quickly decreases to zero.  Therefore, in the
BEC regime, all the large $d$ curves quickly converge and approach the
BEC asymptote. However, in the BCS regime, the detrimental effect of
the mismatch causes $T_c$ to bend back towards stronger interaction in
the low $T$ regime.  (For $k_Fd> 5.4$, one loses superfluidity
completely at $1/k_Fa<0$). For $k_Fd=4$, this fraction remains sizable
as $T_c$ vanishes in the BCS regime; this is the case for which the
Fermi surface mismatch is nearly the least, so that superfluidity is
allowed with such a small mismatch. Except for the $k_Fd=4$ case, for
all other large $d$ cases in Fig.~\ref{fig:fraction}, the fraction
drops to zero as the $T_c$ curves bend back towards BEC and decrease
to 0. This suggests that pairing has to be strong enough so as to pull
all down spin fermions back into the first Brillouin zone, in order to
have a superfluid at zero $T$.

\begin{figure}
  \centerline{\includegraphics[clip,width=3.2in] {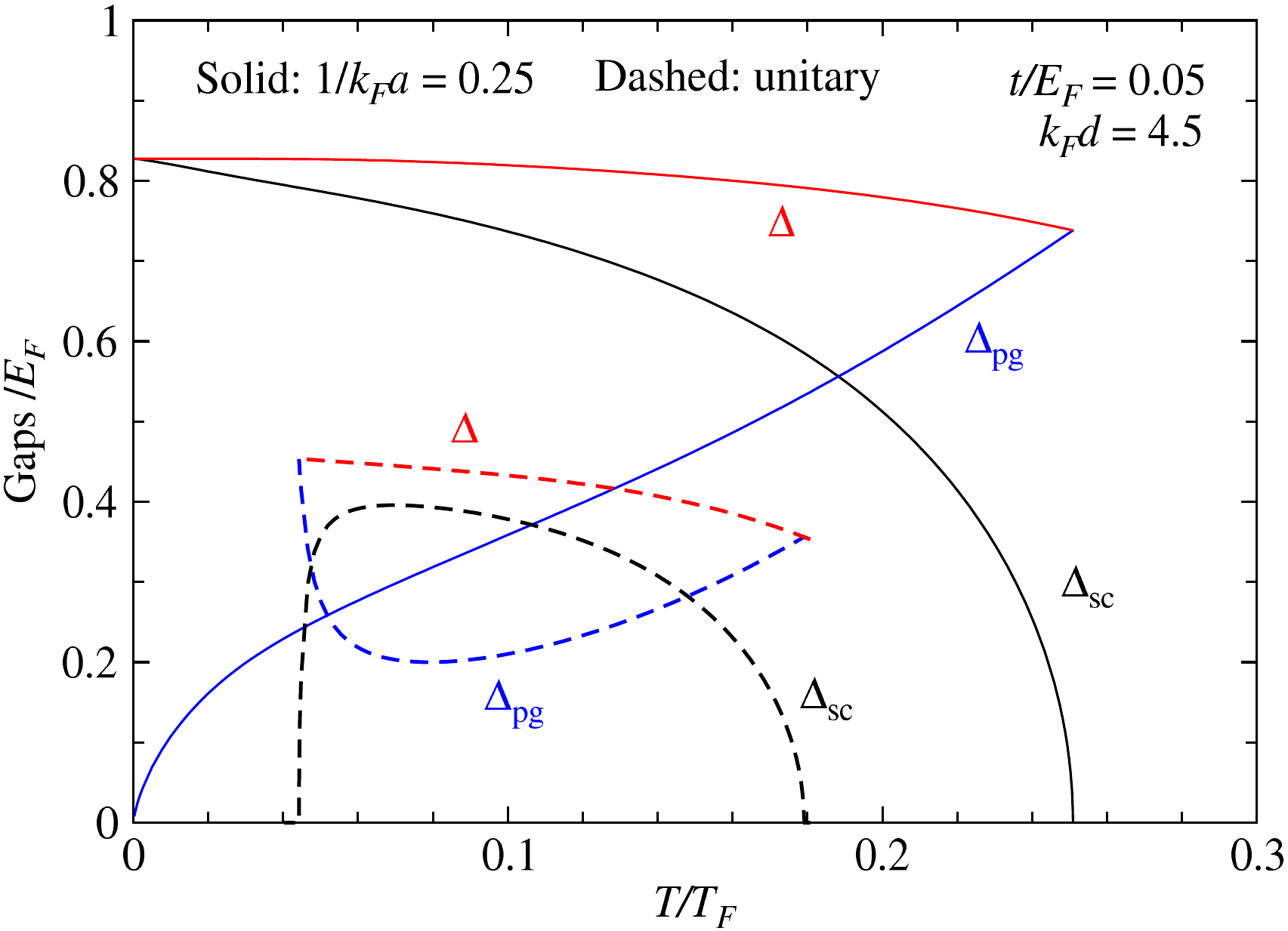}}
  \caption{Typical behaviors of the order parameter $\Delta_{sc}$
    (black) and the gaps $\Delta$ (red) and $\Delta_{pg}$ (blue
    curves), as a function of $T/T_F$ for $1/k_Fa=0.25$ (solid) and
    unitary (dashed lines), representing cases without and with
    intermediate temperature superfluidity, respectively. Here
    $t/E_F = 0.05$ and $k_Fd=4.5$, as labeled. }
\label{fig:gap}
\end{figure}

The back bending of $T_c$ at large $d$ leads to a pronounced
intermediate temperature superfluid behavior. We show in
Fig.~\ref{fig:gap} representative behaviors of the gaps ($\Delta$,
$\Delta_{pg}$) and superfluid order parameter $\Delta_{sc}$ as a
function of $T/T_F$, for the case with (dashed) intermediate
temperature superfluidity, and compare with the case without (solid
lines). For the former case, the order parameter vanishes at both
lower and upper $T_c$'s, sandwiched by pseudogap phases above and
below.

We have also shown in Fig.~\ref{fig:Tc-d} the (yellow shaded) area in
which the $T_c$ solution is unstable against phase separation. In
comparison with the phase diagram of Fermi gases in a simple 3D
continuum in the presence of population imbalance \cite{Chien06}, this
unstable area is very small. We notice that this area exists only for
small $d$ cases. For large $d$, when $T_c$ becomes nonzero, the Fermi
surface mismatch is already alleviated by pairing. 

\begin{figure}
  \centerline{\includegraphics[clip,width=3.4in] {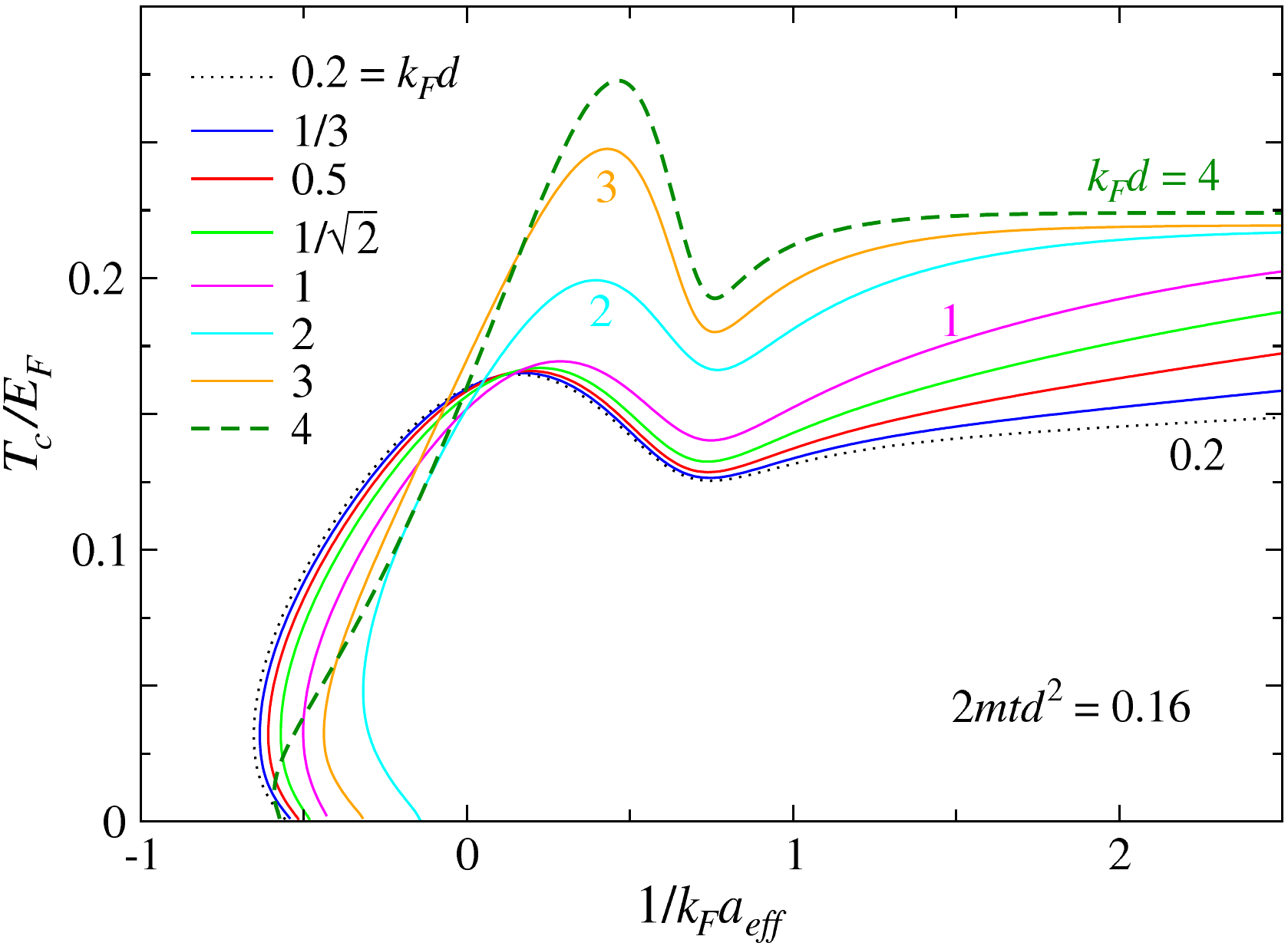}}
  \caption{Behavior of $T_c$ as a function of $1/k_Fa$
    for fixed $2mtd^2 = 0.16$. Except for large $d$, the $T_c$ curves
    are close in the BCS and crossover regimes, while the
    discrepancies become more pronounced in the BEC regime. The
    $k_Fd=4$ case has a good Fermi surface match, exhibiting the least
    frustration on pairing. }
\label{fig:Tc-kd2}
\vspace*{-2ex}
\end{figure}

Finally, we investigate the behavior of $T_c$ with a fixed
$m/m^*_{z,\uparrow} = 2mtd^2$ but different $(t,d)$ combinations.
This corresponds to fixed long wave length effective mass of the
lattice component in the $\hat{z}$ direction. The curves would
collapse to each other should the low $k_z$ part of spin up fermions
dominate the $T_c$ behavior. Shown in Fig.~\ref{fig:Tc-kd2} is a case
with a small $m/m^*_{z,\uparrow} = 0.16$, which is realistic for
experiment. While the curves more or less converge in the fermionic
regime, they separate on the BEC side of the Feshbach resonance. In
the BCS regime, for small $d$, $\pi/d\gg k_F$, therefore, the lattice
effect is not strong. In contrast, in the BEC regime, the BCS
coherence factor $v_\mathbf{k}^2$ (i.e., momentum space pair
occupation number) spreads throughout the entire $k_z$ space, making
the optical lattice effect fully probed.  When $d$ is large, say,
$\pi/d < k_F$, lattice effect will be easily probed even in the BCS
regime, leading to a more pronounced departure.

In summary, we studied the behavior of the entire BCS-BEC crossover at
finite temperature in mixed-dimensional Fermi gases using a pairing
fluctuation theory. We found that tunable mixed dimensionality can
create large Fermi surface mismatches. The $T_c$ solutions bear
similarity with simple population imbalance Fermi gases in a 3D
continuum, but with some distinct features. While intermediate
temperature superfluidity also exists, reentrant superfluid behavior
with a pair density wave ground state in between emerges at small
$d$. Unlike an pure optical lattice case, $T_c$ approaches a constant
asymptote in the deep BEC regime.  With modern techniques, these
predictions can be tested experimentally.

We thank useful discussions with Brandon Anderson, Rufus Boyack,
A.J. Leggett, K. Levin, and Bo Yan. This work is supported by NSF of
China (Grants No. 10974173 and No. 11274267), the National Basic
Research Program of China (Grants No. 2011CB921303 and
No. 2012CB927404), NSF of Zhejiang Province of China (Grant No.
LZ13A040001).


\begin{thebibliography}{21}%
\makeatletter
\providecommand \@ifxundefined [1]{%
 \@ifx{#1\undefined}
}%
\providecommand \@ifnum [1]{%
 \ifnum #1\expandafter \@firstoftwo
 \else \expandafter \@secondoftwo
 \fi
}%
\providecommand \@ifx [1]{%
 \ifx #1\expandafter \@firstoftwo
 \else \expandafter \@secondoftwo
 \fi
}%
\providecommand \natexlab [1]{#1}%
\providecommand \enquote  [1]{``#1''}%
\providecommand \bibnamefont  [1]{#1}%
\providecommand \bibfnamefont [1]{#1}%
\providecommand \citenamefont [1]{#1}%
\providecommand \href@noop [0]{\@secondoftwo}%
\providecommand \href [0]{\begingroup \@sanitize@url \@href}%
\providecommand \@href[1]{\@@startlink{#1}\@@href}%
\providecommand \@@href[1]{\endgroup#1\@@endlink}%
\providecommand \@sanitize@url [0]{\catcode `\\12\catcode `\$12\catcode
  `\&12\catcode `\#12\catcode `\^12\catcode `\_12\catcode `\%12\relax}%
\providecommand \@@startlink[1]{}%
\providecommand \@@endlink[0]{}%
\providecommand \url  [0]{\begingroup\@sanitize@url \@url }%
\providecommand \@url [1]{\endgroup\@href {#1}{\urlprefix }}%
\providecommand \urlprefix  [0]{URL }%
\providecommand \Eprint [0]{\href }%
\providecommand \doibase [0]{http://dx.doi.org/}%
\providecommand \selectlanguage [0]{\@gobble}%
\providecommand \bibinfo  [0]{\@secondoftwo}%
\providecommand \bibfield  [0]{\@secondoftwo}%
\providecommand \translation [1]{[#1]}%
\providecommand \BibitemOpen [0]{}%
\providecommand \bibitemStop [0]{}%
\providecommand \bibitemNoStop [0]{.\EOS\space}%
\providecommand \EOS [0]{\spacefactor3000\relax}%
\providecommand \BibitemShut  [1]{\csname bibitem#1\endcsname}%
\let\auto@bib@innerbib\@empty
\bibitem [{\citenamefont {Chen}\ \emph {et~al.}(2005)\citenamefont {Chen},
  \citenamefont {Stajic}, \citenamefont {Tan},\ and\ \citenamefont
  {Levin}}]{Review}%
  \BibitemOpen
  \bibfield  {author} {\bibinfo {author} {\bibfnamefont {Q~J}\ \bibnamefont
  {Chen}}, \bibinfo {author} {\bibfnamefont {J}~\bibnamefont {Stajic}},
  \bibinfo {author} {\bibfnamefont {S~N}\ \bibnamefont {Tan}}, \ and\ \bibinfo
  {author} {\bibfnamefont {K}~\bibnamefont {Levin}},\ }\bibfield  {title}
  {\enquote {\bibinfo {title} {{BCS-BEC} crossover: From high temperature
  superconductors to ultracold superfluids},}\ }\href@noop {} {\bibfield
  {journal} {\bibinfo  {journal} {Phys. Rep.}\ }\textbf {\bibinfo {volume}
  {412}},\ \bibinfo {pages} {1--88} (\bibinfo {year} {2005})}\BibitemShut
  {NoStop}%
\bibitem [{\citenamefont {Bloch}\ \emph {et~al.}(2008)\citenamefont {Bloch},
  \citenamefont {Dalibard},\ and\ \citenamefont {Zwerger}}]{Bloch_RMP}%
  \BibitemOpen
  \bibfield  {author} {\bibinfo {author} {\bibfnamefont {Immanuel}\
  \bibnamefont {Bloch}}, \bibinfo {author} {\bibfnamefont {Jean}\ \bibnamefont
  {Dalibard}}, \ and\ \bibinfo {author} {\bibfnamefont {Wilhelm}\ \bibnamefont
  {Zwerger}},\ }\bibfield  {title} {\enquote {\bibinfo {title} {Many-body
  physics with ultracold gases},}\ }\href {\doibase 10.1103/RevModPhys.80.885}
  {\bibfield  {journal} {\bibinfo  {journal} {Rev. Mod. Phys.}\ }\textbf
  {\bibinfo {volume} {80}},\ \bibinfo {pages} {885--964} (\bibinfo {year}
  {2008})}\BibitemShut {NoStop}%
\bibitem [{\citenamefont {Bloch}(2005)}]{Bloch05NP}%
  \BibitemOpen
  \bibfield  {author} {\bibinfo {author} {\bibfnamefont {Immanuel}\
  \bibnamefont {Bloch}},\ }\bibfield  {title} {\enquote {\bibinfo {title}
  {Ultracold quantum gases in optical lattices},}\ }\href {\doibase
  10.1038/nphys138} {\bibfield  {journal} {\bibinfo  {journal} {Nature
  Physics}\ }\textbf {\bibinfo {volume} {1}},\ \bibinfo {pages} {23--30}
  (\bibinfo {year} {2005})}\BibitemShut {NoStop}%
\bibitem [{\citenamefont {Wu}\ \emph {et~al.}(2015)\citenamefont {Wu},
  \citenamefont {Anderson}, \citenamefont {Boyack},\ and\ \citenamefont
  {Levin}}]{CWu15PRL}%
  \BibitemOpen
  \bibfield  {author} {\bibinfo {author} {\bibfnamefont {Chien-Te}\
  \bibnamefont {Wu}}, \bibinfo {author} {\bibfnamefont {Brandon~M.}\
  \bibnamefont {Anderson}}, \bibinfo {author} {\bibfnamefont {Rufus}\
  \bibnamefont {Boyack}}, \ and\ \bibinfo {author} {\bibfnamefont
  {K.}~\bibnamefont {Levin}},\ }\bibfield  {title} {\enquote {\bibinfo {title}
  {Quasicondensation in two-dimensional Fermi gases},}\ }\href {\doibase
  10.1103/PhysRevLett.115.240401} {\bibfield  {journal} {\bibinfo  {journal}
  {Phys. Rev. Lett.}\ }\textbf {\bibinfo {volume} {115}},\ \bibinfo {pages}
  {240401} (\bibinfo {year} {2015})}\BibitemShut {NoStop}%
\bibitem [{\citenamefont {Wu}\ \emph {et~al.}(2016)\citenamefont {Wu},
  \citenamefont {Anderson}, \citenamefont {Boyack},\ and\ \citenamefont
  {Levin}}]{CWu16}%
  \BibitemOpen
  \bibfield  {author} {\bibinfo {author} {\bibfnamefont {Chien-Te}\
  \bibnamefont {Wu}}, \bibinfo {author} {\bibfnamefont {Brandon~M.}\
  \bibnamefont {Anderson}}, \bibinfo {author} {\bibfnamefont {Rufus}\
  \bibnamefont {Boyack}}, \ and\ \bibinfo {author} {\bibfnamefont
  {K.}~\bibnamefont {Levin}},\ }\bibfield  {title} {\enquote {\bibinfo {title}
  {Two-dimensional spin-imbalanced {F}ermi gases at non-zero temperature:
  {P}hase separation of a non-condensate},}\ }\href@noop {} {\bibfield
  {journal} {\bibinfo  {journal} {Phys. Rev. A}\ }\textbf {\bibinfo {volume}
  {94}},\ \bibinfo {pages} {033604} (\bibinfo {year} {2016})}\BibitemShut
  {NoStop}%
\bibitem [{\citenamefont {Nishida}\ and\ \citenamefont
  {Tan}(2008)}]{Nishida08PRL}%
  \BibitemOpen
  \bibfield  {author} {\bibinfo {author} {\bibfnamefont {Yusuke}\ \bibnamefont
  {Nishida}}\ and\ \bibinfo {author} {\bibfnamefont {Shina}\ \bibnamefont
  {Tan}},\ }\bibfield  {title} {\enquote {\bibinfo {title} {Universal Fermi
  gases in mixed dimensions},}\ }\href {\doibase
  10.1103/PhysRevLett.101.170401} {\bibfield  {journal} {\bibinfo  {journal}
  {Phys. Rev. Lett.}\ }\textbf {\bibinfo {volume} {101}},\ \bibinfo {pages}
  {170401} (\bibinfo {year} {2008})}\BibitemShut {NoStop}%
\bibitem [{\citenamefont {Nishida}(2010)}]{Nishida10PRA}%
  \BibitemOpen
  \bibfield  {author} {\bibinfo {author} {\bibfnamefont {Yusuke}\ \bibnamefont
  {Nishida}},\ }\bibfield  {title} {\enquote {\bibinfo {title} {Phases of a
  bilayer Fermi gas},}\ }\href {\doibase 10.1103/PhysRevA.82.011605} {\bibfield
   {journal} {\bibinfo  {journal} {Phys. Rev. A}\ }\textbf {\bibinfo {volume}
  {82}},\ \bibinfo {pages} {011605} (\bibinfo {year} {2010})}\BibitemShut
  {NoStop}%
\bibitem [{\citenamefont {Lamporesi}\ \emph {et~al.}(2010)\citenamefont
  {Lamporesi}, \citenamefont {Catani}, \citenamefont {Barontini}, \citenamefont
  {Nishida}, \citenamefont {Inguscio},\ and\ \citenamefont
  {Minardi}}]{Lamporesi10PRL}%
  \BibitemOpen
  \bibfield  {author} {\bibinfo {author} {\bibfnamefont {G.}~\bibnamefont
  {Lamporesi}}, \bibinfo {author} {\bibfnamefont {J.}~\bibnamefont {Catani}},
  \bibinfo {author} {\bibfnamefont {G.}~\bibnamefont {Barontini}}, \bibinfo
  {author} {\bibfnamefont {Y.}~\bibnamefont {Nishida}}, \bibinfo {author}
  {\bibfnamefont {M.}~\bibnamefont {Inguscio}}, \ and\ \bibinfo {author}
  {\bibfnamefont {F.}~\bibnamefont {Minardi}},\ }\bibfield  {title} {\enquote
  {\bibinfo {title} {Scattering in mixed dimensions with ultracold gases},}\
  }\href {\doibase 10.1103/PhysRevLett.104.153202} {\bibfield  {journal}
  {\bibinfo  {journal} {Phys. Rev. Lett.}\ }\textbf {\bibinfo {volume} {104}},\
  \bibinfo {pages} {153202} (\bibinfo {year} {2010})}\BibitemShut {NoStop}%
\bibitem [{\citenamefont {Iskin}\ and\ \citenamefont {Suba\ifmmode
  \mbox{\c{s}}\else \c{s}\fi{}\ifmmode \imath \else~\i
  \fi{}}(2010)}]{Iskin10PRA}%
  \BibitemOpen
  \bibfield  {author} {\bibinfo {author} {\bibfnamefont {M.}~\bibnamefont
  {Iskin}}\ and\ \bibinfo {author} {\bibfnamefont {A.~L.}\ \bibnamefont
  {Suba\ifmmode \mbox{\c{s}}\else \c{s}\fi{}\ifmmode \imath \else~\i \fi{}}},\
  }\bibfield  {title} {\enquote {\bibinfo {title} {Cooper pairing and BCS-BEC
  evolution in mixed-dimensional Fermi gases},}\ }\href {\doibase
  10.1103/PhysRevA.82.063628} {\bibfield  {journal} {\bibinfo  {journal} {Phys.
  Rev. A}\ }\textbf {\bibinfo {volume} {82}},\ \bibinfo {pages} {063628}
  (\bibinfo {year} {2010})}\BibitemShut {NoStop}%
\bibitem [{\citenamefont {Yang}\ \emph {et~al.}(2011)\citenamefont {Yang},
  \citenamefont {Huang},\ and\ \citenamefont {Wan}}]{XYang11EPJB}%
  \BibitemOpen
  \bibfield  {author} {\bibinfo {author} {\bibfnamefont {X.~S.}\ \bibnamefont
  {Yang}}, \bibinfo {author} {\bibfnamefont {B.~B.}\ \bibnamefont {Huang}}, \
  and\ \bibinfo {author} {\bibfnamefont {S.~L.}\ \bibnamefont {Wan}},\
  }\bibfield  {title} {\enquote {\bibinfo {title} {BCS-BEC crossover in
  mix-dimensional Fermi gases},}\ }\href {\doibase 10.1140/epjb/e2011-20354-0}
  {\bibfield  {journal} {\bibinfo  {journal} {Eur. Phys. J. B}\ }\textbf
  {\bibinfo {volume} {83}},\ \bibinfo {pages} {445--450} (\bibinfo {year}
  {2011})}\BibitemShut {NoStop}%
\bibitem [{\citenamefont {Fulde}\ and\ \citenamefont {Ferrell}(1964)}]{FF}%
  \BibitemOpen
  \bibfield  {author} {\bibinfo {author} {\bibfnamefont {Peter}\ \bibnamefont
  {Fulde}}\ and\ \bibinfo {author} {\bibfnamefont {Richard~A.}\ \bibnamefont
  {Ferrell}},\ }\bibfield  {title} {\enquote {\bibinfo {title}
  {Superconductivity in a strong spin-exchange field},}\ }\href {\doibase
  10.1103/PhysRev.135.A550} {\bibfield  {journal} {\bibinfo  {journal} {Phys.
  Rev.}\ }\textbf {\bibinfo {volume} {135}},\ \bibinfo {pages} {A550--A563}
  (\bibinfo {year} {1964})}\BibitemShut {NoStop}%
\bibitem [{\citenamefont {Larkin}\ and\ \citenamefont
  {Ovchinnikov}(1965)}]{LO}%
  \BibitemOpen
  \bibfield  {author} {\bibinfo {author} {\bibfnamefont {A.~I.}\ \bibnamefont
  {Larkin}}\ and\ \bibinfo {author} {\bibfnamefont {Y.~N.}\ \bibnamefont
  {Ovchinnikov}},\ }\bibfield  {title} {\enquote {\bibinfo {title}
  {Inhomogeneous state of superconductors},}\ }\href@noop {} {\bibfield
  {journal} {\bibinfo  {journal} {Sov. Phys. JETP}\ }\textbf {\bibinfo {volume}
  {20}},\ \bibinfo {pages} {762--769} (\bibinfo {year} {1965})},\ \bibinfo
  {note} {[Zh. Eksp. Teor. Fiz. \textbf{47}, 1136 (1964)]}\BibitemShut
  {NoStop}%
\bibitem [{\citenamefont {Chen}\ \emph {et~al.}(1998)\citenamefont {Chen},
  \citenamefont {Kosztin}, \citenamefont {Jank\'o},\ and\ \citenamefont
  {Levin}}]{Chen2}%
  \BibitemOpen
  \bibfield  {author} {\bibinfo {author} {\bibfnamefont {Q~J}\ \bibnamefont
  {Chen}}, \bibinfo {author} {\bibfnamefont {I}~\bibnamefont {Kosztin}},
  \bibinfo {author} {\bibfnamefont {B}~\bibnamefont {Jank\'o}}, \ and\ \bibinfo
  {author} {\bibfnamefont {K}~\bibnamefont {Levin}},\ }\bibfield  {title}
  {\enquote {\bibinfo {title} {Pairing fluctuation theory of superconducting
  properties in underdoped to overdoped cuprates.}}\ }\href@noop {} {\bibfield
  {journal} {\bibinfo  {journal} {Phys. Rev. Lett.}\ }\textbf {\bibinfo
  {volume} {81}},\ \bibinfo {pages} {4708--11} (\bibinfo {year}
  {1998})}\BibitemShut {NoStop}%
\bibitem [{\citenamefont {Chen}\ and\ \citenamefont {Wang}(2014)}]{FrontPhys}%
  \BibitemOpen
  \bibfield  {author} {\bibinfo {author} {\bibfnamefont {Q.~J.}\ \bibnamefont
  {Chen}}\ and\ \bibinfo {author} {\bibfnamefont {J~B}\ \bibnamefont {Wang}},\
  }\bibfield  {title} {\enquote {\bibinfo {title} {Pseudogap phenomena in
  ultracold atomic Fermi gases},}\ }\href@noop {} {\bibfield  {journal}
  {\bibinfo  {journal} {Front. Phys.}\ }\textbf {\bibinfo {volume} {9}},\
  \bibinfo {pages} {539--570} (\bibinfo {year} {2014})}\BibitemShut {NoStop}%
\bibitem [{\citenamefont {Chien}\ \emph {et~al.}(2006)\citenamefont {Chien},
  \citenamefont {Chen}, \citenamefont {He},\ and\ \citenamefont
  {Levin}}]{Chien06}%
  \BibitemOpen
  \bibfield  {author} {\bibinfo {author} {\bibfnamefont {C~C}\ \bibnamefont
  {Chien}}, \bibinfo {author} {\bibfnamefont {Q~J}\ \bibnamefont {Chen}},
  \bibinfo {author} {\bibfnamefont {Y}~\bibnamefont {He}}, \ and\ \bibinfo
  {author} {\bibfnamefont {K}~\bibnamefont {Levin}},\ }\bibfield  {title}
  {\enquote {\bibinfo {title} {Intermediate temperature superfluidity in a
  Fermi gas with population imbalance},}\ }\href@noop {} {\bibfield  {journal}
  {\bibinfo  {journal} {Phys. Rev. Lett.}\ }\textbf {\bibinfo {volume} {97}},\
  \bibinfo {pages} {090402} (\bibinfo {year} {2006})}\BibitemShut {NoStop}%
\bibitem [{\citenamefont {Guo}\ \emph {et~al.}(2009)\citenamefont {Guo},
  \citenamefont {Chien}, \citenamefont {Chen}, \citenamefont {He},\ and\
  \citenamefont {Levin}}]{Guo2009PRA}%
  \BibitemOpen
  \bibfield  {author} {\bibinfo {author} {\bibfnamefont {Hao}\ \bibnamefont
  {Guo}}, \bibinfo {author} {\bibfnamefont {Chih-Chun}\ \bibnamefont {Chien}},
  \bibinfo {author} {\bibfnamefont {Q~J}\ \bibnamefont {Chen}}, \bibinfo
  {author} {\bibfnamefont {Yan}\ \bibnamefont {He}}, \ and\ \bibinfo {author}
  {\bibfnamefont {K.}~\bibnamefont {Levin}},\ }\bibfield  {title} {\enquote
  {\bibinfo {title} {Finite-temperature behavior of an interspecies fermionic
  superfluid with population imbalance},}\ }\href {\doibase
  10.1103/PhysRevA.80.011601} {\bibfield  {journal} {\bibinfo  {journal} {Phys.
  Rev. A}\ }\textbf {\bibinfo {volume} {80}},\ \bibinfo {pages} {011601(R)}
  (\bibinfo {year} {2009})}\BibitemShut {NoStop}%
\bibitem [{\citenamefont {Micnas}\ \emph {et~al.}(1990)\citenamefont {Micnas},
  \citenamefont {Ranninger},\ and\ \citenamefont {Robaszkiewicz}}]{Micnas}%
  \BibitemOpen
  \bibfield  {author} {\bibinfo {author} {\bibfnamefont {R}~\bibnamefont
  {Micnas}}, \bibinfo {author} {\bibfnamefont {J}~\bibnamefont {Ranninger}}, \
  and\ \bibinfo {author} {\bibfnamefont {S}~\bibnamefont {Robaszkiewicz}},\
  }\bibfield  {title} {\enquote {\bibinfo {title} {Superconductivity in
  narrow-band systems with local nonretarded attractive interactions.}}\
  }\href@noop {} {\bibfield  {journal} {\bibinfo  {journal} {Rev. Mod. Phys.}\
  }\textbf {\bibinfo {volume} {62}},\ \bibinfo {pages} {113--171} (\bibinfo
  {year} {1990})}\BibitemShut {NoStop}%
\bibitem [{\citenamefont {Pao}\ \emph {et~al.}(2006)\citenamefont {Pao},
  \citenamefont {Wu},\ and\ \citenamefont {Yip}}]{PWY05}%
  \BibitemOpen
  \bibfield  {author} {\bibinfo {author} {\bibfnamefont {C.-H.}\ \bibnamefont
  {Pao}}, \bibinfo {author} {\bibfnamefont {Shin-Tza}\ \bibnamefont {Wu}}, \
  and\ \bibinfo {author} {\bibfnamefont {S.-K.}\ \bibnamefont {Yip}},\
  }\bibfield  {title} {\enquote {\bibinfo {title} {Superfluid stability in the
  BEC-BCS crossover},}\ }\href {\doibase 10.1103/PhysRevB.73.132506} {\bibfield
   {journal} {\bibinfo  {journal} {Phys. Rev. B}\ }\textbf {\bibinfo {volume}
  {73}},\ \bibinfo {pages} {132506} (\bibinfo {year} {2006})}\BibitemShut
  {NoStop}%
\bibitem [{\citenamefont {Chen}\ \emph {et~al.}(2006)\citenamefont {Chen},
  \citenamefont {He}, \citenamefont {Chien},\ and\ \citenamefont
  {Levin}}]{Stability}%
  \BibitemOpen
  \bibfield  {author} {\bibinfo {author} {\bibfnamefont {Q~J}\ \bibnamefont
  {Chen}}, \bibinfo {author} {\bibfnamefont {Yan}\ \bibnamefont {He}}, \bibinfo
  {author} {\bibfnamefont {Chih-Chun}\ \bibnamefont {Chien}}, \ and\ \bibinfo
  {author} {\bibfnamefont {K.}~\bibnamefont {Levin}},\ }\bibfield  {title}
  {\enquote {\bibinfo {title} {Stability conditions and phase diagrams for
  two-component Fermi gases with population imbalance},}\ }\href {\doibase
  10.1103/PhysRevA.74.063603} {\bibfield  {journal} {\bibinfo  {journal} {Phys.
  Rev. A}\ }\textbf {\bibinfo {volume} {74}},\ \bibinfo {pages} {063603}
  (\bibinfo {year} {2006})}\BibitemShut {NoStop}%
\bibitem [{\citenamefont {Wang}\ \emph {et~al.}(2013)\citenamefont {Wang},
  \citenamefont {Guo},\ and\ \citenamefont {Chen}}]{wang13pra}%
  \BibitemOpen
  \bibfield  {author} {\bibinfo {author} {\bibfnamefont {J~B}\ \bibnamefont
  {Wang}}, \bibinfo {author} {\bibfnamefont {Hao}\ \bibnamefont {Guo}}, \ and\
  \bibinfo {author} {\bibfnamefont {Q~J}\ \bibnamefont {Chen}},\ }\bibfield
  {title} {\enquote {\bibinfo {title} {Exotic phase separation and phase
  diagrams of a Fermi-Fermi mixture in a trap at finite temperature},}\ }\href
  {\doibase 10.1103/PhysRevA.87.041601} {\bibfield  {journal} {\bibinfo
  {journal} {Phys. Rev. A}\ }\textbf {\bibinfo {volume} {87}},\ \bibinfo
  {pages} {041601(R)} (\bibinfo {year} {2013})}\BibitemShut {NoStop}%
\bibitem [{\citenamefont {Che}\ \emph {et~al.}(2016)\citenamefont {Che},
  \citenamefont {Wang},\ and\ \citenamefont {Chen}}]{ChePolar}%
  \BibitemOpen
  \bibfield  {author} {\bibinfo {author} {\bibfnamefont {Yanming}\
  \bibnamefont {Che}}, \bibinfo {author} {\bibfnamefont {Jibiao}\ \bibnamefont
  {Wang}}, \ and\ \bibinfo {author} {\bibfnamefont {Q.~J.}\ \bibnamefont
  {Chen}},\ }\bibfield  {title} {\enquote {\bibinfo {title} {Reentrant
  superfluidity and pair density wave in single component dipolar Fermi
  gases},}\ }\href@noop {} {\bibfield  {journal} {\bibinfo  {journal} {\pra}\
  }\textbf {\bibinfo {volume} {93}},\ \bibinfo {pages} {063611} (\bibinfo
  {year} {2016})}\BibitemShut {NoStop}%
\end{thebibliography}
%

\end{document}


\title{Supplementary Information\\
Exotic superfluidity and pairing phenomena in atomic Fermi
  gases in mixed dimensions}

\author{ Leifeng Zhang }
\author{Yanming Che }
\affiliation{Department of Physics and Zhejiang Institute of Modern
  Physics, Zhejiang University, Hangzhou, Zhejiang 310027, China}
\affiliation{Synergetic Innovation Center of Quantum Information and
  Quantum Physics, Hefei, Anhui 230026, China} 

\author{Jibiao Wang }
\affiliation{Department of Physics and Zhejiang Institute of Modern
  Physics, Zhejiang University, Hangzhou, Zhejiang 310027, China}
\affiliation{Synergetic Innovation Center of Quantum Information and
  Quantum Physics, Hefei, Anhui 230026, China} 
\affiliation{TianQin Research Center \& School of Physics and Astronomy, Sun Yat-Sen University (Zhuhai Campus), Zhuhai, Guangdong 519082, China} 

\author{Qijin Chen}
\email[Corresponding author: ]{qchen@uchicago.edu}
\affiliation{Department of Physics and Zhejiang Institute of Modern
  Physics, Zhejiang University, Hangzhou, Zhejiang 310027, China}
\affiliation{Synergetic Innovation Center of Quantum Information and
  Quantum Physics, Hefei, Anhui 230026, China} 
\affiliation{James Franck Institute, University of Chicago, Chicago, Illinois
  60637, USA}

\date{\today}

\begin{abstract}
Here we present extra plots which may help with the understanding of the main text.
\end{abstract}


\maketitle

\section{Superfluid transition $T_c$ as a function of $1/k_Fa$ for $t/E_F=1$}

\begin{figure}[t]
  \centerline{\includegraphics[clip,width=3.2in] {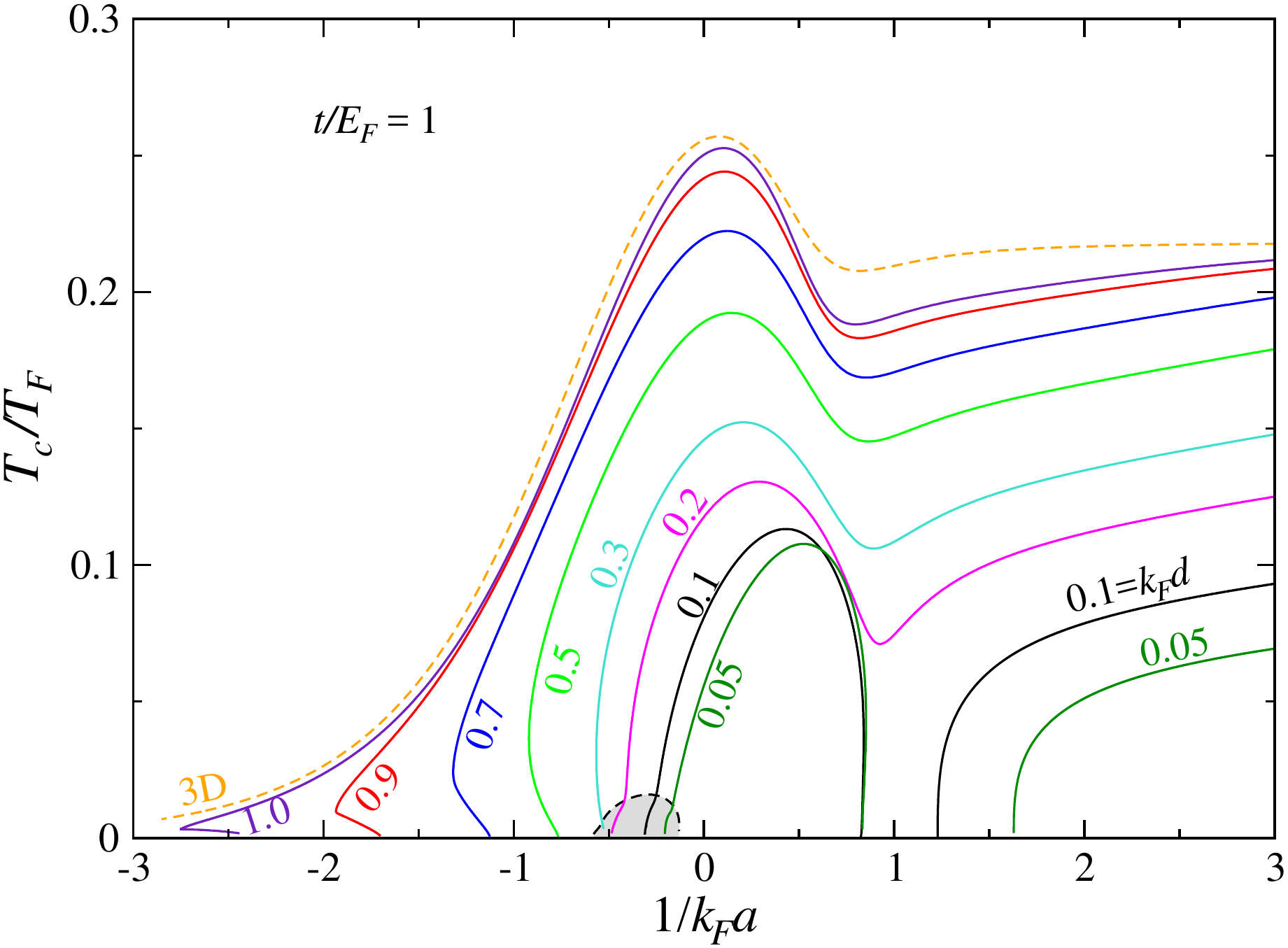}}
  \caption{ Behavior of $T_c$ as functions of $1/k_Fa$ at fixed
    $t/E_F = 1$, but for different value of $k_Fd \le 1$, as
    labeled. The $T_c$ solution in shaded regions is unstable against
    phase separation.}
\label{fig:Tc-d}
\end{figure}

Shown in Fig.~\ref{fig:Tc-d} is $T_c$ as a function of $1/k_Fa$ for
$t/E_F=1$, with a series of values of $k_Fd$, as labeled. As a basis
for comparison, we also included the $T_c$ curve from a simple
isotropic 3D Fermi gas, labeled ``3D''. For this large $t=E_F$, the
best Fermi surface match occurs near $k_Fd=1$. Here we only show the
curves with $k_Fd<1$, which do not intersect the 3D curve. The $T_c$
curve splits for small $d$, giving way to pair density wave ground
states. In the shaded area, the system is unstable at
$T_c$. Intermediate temperature superfluid exists for $k_Fd \ge 0.3$.

\section{Effects of a band dispersion for pairs}

\begin{figure}[b]
  \centerline{\includegraphics[clip,width=3.2in] {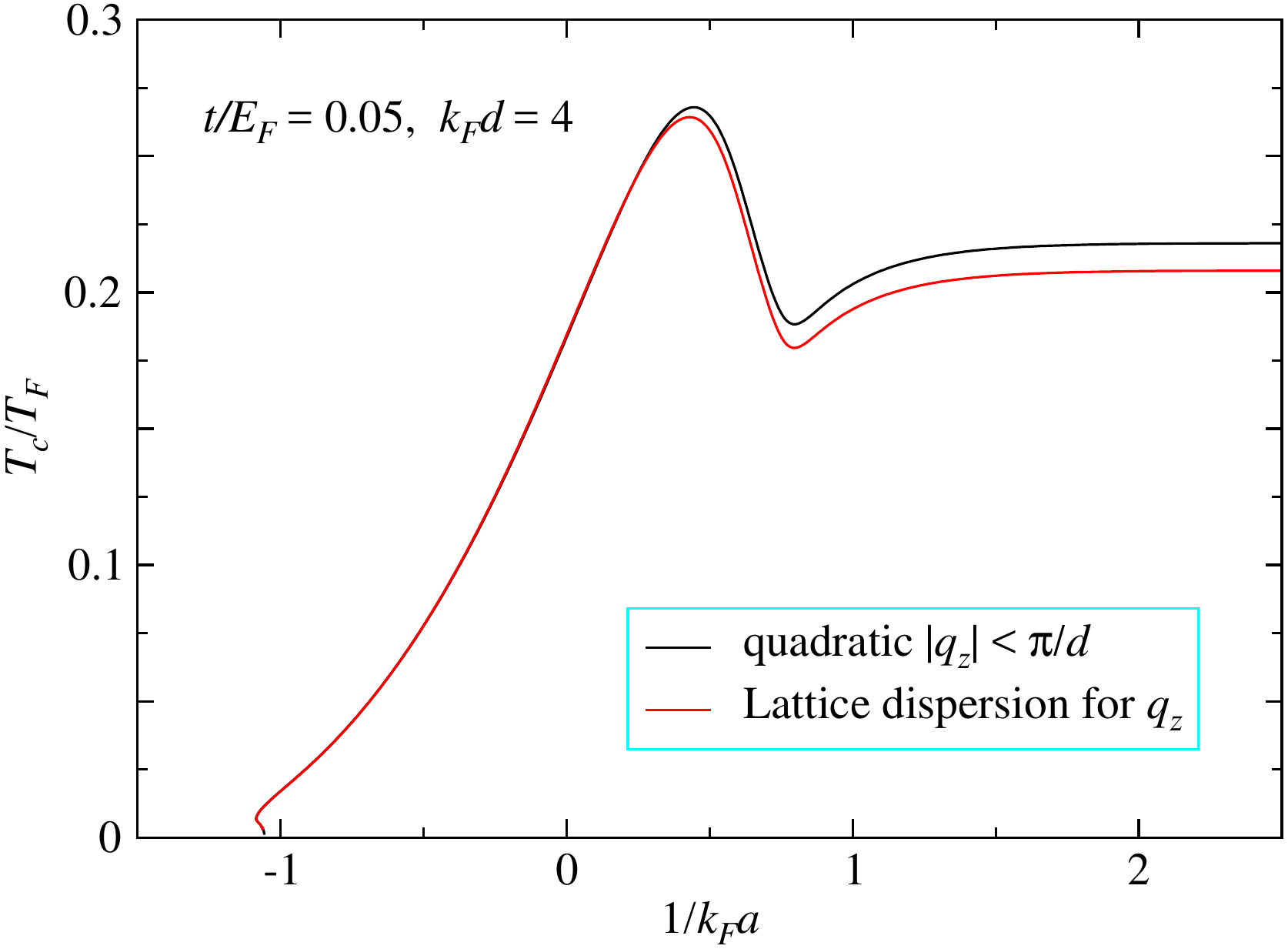}}
  \caption{Comparison between two $T_c$ solutions as a function of
    $1/k_Fa$ using a parapolic dispersion (black) and a band
    dispersion (red) for the $q_z$ contribution of the pair. Here
    $t/E_F = 0.05$ and $k_Fd=4$, as labeled. }
\label{fig:qz-band}
\end{figure}

To check the effect of a band dispersion for the pairs on $T_c$, we
performed $T_c$ calculations using both parabolic and band dispersions
for the $\hat{z}$ direction of spin-up fermions. The result is shown
in Fig.~\ref{fig:qz-band}, for $t/E_F=0.05$ and $k_Fd=4$. It is
evident that the two curves overlap with each other for $1/k_Fa<0$,
and only a minor quantitative difference arises in the BEC regime.

